\documentclass[12pt]{article}
\usepackage{epsfig}
\usepackage{cite}
\usepackage{color}
\usepackage{amssymb}
\usepackage{float}
\usepackage{cite}
\def  \bcen   {\begin{center}}
\def  \ecen   {\end{center}}
\def  \beq    {\begin{equation}}
\def  \eeq    {\end{equation}}
\def  \beqa   {\begin{eqnarray}}
\def  \eeqa   {\end{eqnarray}}

\def\bea{\begin{eqnarray}}
\def\eea{\end{eqnarray}}
\def \lumi        {fb$^{-1}$ }
\def \siggaga     {$\sigma^{\gamma\gamma}$ }
\def \htogaga     {$H \to \gamma \gamma$ }
\def \rgaga       {$R_{\gamma\gamma}$ }
\def \redbullet   {\textcolor{red}{\large \textbullet}}
\def \bluebullet  {\textcolor{blue}{\large \textbullet}}

\begin{document}
\def\thesection {\Roman{section}}
\def\thesubsection {\Alph{subsection}}
\renewcommand{\thefootnote}{\fnsymbol{footnote}}
\vspace*{1cm}
\bcen
{\Large \bf \boldmath 
\htogaga in Inert Higgs Doublet Model  
}
\vskip 2cm
{\bf
Abdesslam Arhrib $^{1,2}$\footnote{aarhrib@ictp.it},
Rachid Benbrik $^{2,3,4}$\footnote{rbenbrik@ictp.it}, 
Naveen Gaur $^5$\footnote{gaur.nav@gmail.com},
}
\vskip 1cm
{\sl
$^1$ Facult\'e des Sciences et Techniques, B.P 416 Tangier, Morocco.\\
$^{2}$LPHEA, FSSM, Cadi Ayyad University, B.P. 2390, Marrakesh, Morocco.\\
$^{3}$Instituto de F\'isica de Cantabria (CSIC-UC), Santander, Spain.\\
$^{4}$Facult\'e Polydisciplinaire de Safi, Sidi Bouzid B.P 4162, 46000 Safi,
Morocco.\\
$^5$ Department of Physics, Dyal Singh College 
(University of Delhi), Lodi Road \\
New Delhi - 110003, India.
}
\ecen
\vskip 1.5cm

\begin{abstract}
\noindent 

Motivated by the recent result reported from LHC on the 
 di-photon search for a Standard Model (SM) Higgs-like boson.
 We discuss the implications of this possible signal in the framework of the
 Inert Higgs Doublet Model (IHDM), taking into account previous limits from
 Higgs searches at LEP, the Tevatron and the LHC as well as constraints from
 unitarity, vacuum stability and electroweak precision tests.
We show that the charged Higgs contributions can
interfere constructively or destructively with the W gauge 
bosons loops leading to enhancement or suppression of the di-photon rate 
with respect to SM rate.
We show also that the invisible decay of the Higgs, if open, 
could affect the total width of the SM Higgs boson 
and therefore suppress the di-photon rate.
\end{abstract}

\vfill\eject

\renewcommand{\thefootnote}{\arabic{footnote}}
\setcounter{footnote}{0}     

\section{Introduction \label{section:1}}

LHC in $p p$ collision at 7 TeV has already delivered a integrated
luminosity of more than 5 fb$^{-1}$. Based on this delivered
integrated luminosity recently ATLAS \cite{atlas} and CMS \cite{cms} have
presented  their combined and updated results of SM Higgs boson
searches. Both the collaborations attempted to search for the SM Higgs
boson in mass range 110-600 GeV, the main channels used by them for
the analysis are :
\begin{itemize}
\item{} {\bf ATLAS \cite{atlas} } $H \to Z Z^* \to 4 \ell$ and
\htogaga with full data set of 4.8 \lumi and 4.9 
\lumi respectively. Update of $H \to W W^* \to \ell \nu \ell \nu$, $H
\to Z Z^* \to 2 \ell 2 \nu $, $H \to Z Z^* \to 2 \ell 2 q $ with 2.1
\lumi. They reported an excess of 
events around the Higgs mass of 126-127 GeV with the maximum local
significance level of 2.6 $\sigma$. 
\item{} {\bf CMS \cite{cms} :} \htogaga, $H \to b b$, $H \to Z Z^* \to
4 \ell$, $H \to 2 \ell 2 \tau$ at 4.7 \lumi and $H \to \tau \tau$, $H
\to W W^* \to 2 \ell 2 \nu$, $H \to Z Z^* \to 2 \ell 2 \nu$, $H \to Z
Z^* \to 2 \ell 2 q$ at 4.6 \lumi. They reported a local significance
of 2.4 $\sigma$ around the Higgs mass of 124 GeV. 
\end{itemize}
Note that both CMS and ATLAS report some  excess but with lower 
 statistical significance, in the $WW^*$ and $ZZ^*$ channels.
Moreover, from the di-photon channel, 
ATLAS and CMS have excluded a SM Higgs in a small 
portions of this mass range, $114$--$115$~GeV for ATLAS and 
$127$--$131$ GeV for CMS, at the $95 \%$C.L.\\
With $4.9 {\rm fb}^{-1}$ datasets
using the combined channels, both ATLAS by CMS, 
 have narrowed further down the mass window
for a light SM Higgs, excluding respectively the mass 
ranges $131$--$237$ GeV and $251$--$453$ GeV \cite{atlas}, 
and $127$--$600$ GeV \cite{cms} at the  $95 \%$C.L.

The effective cross-section of di-photon ($\gamma \gamma$) mode can
 be estimated by inclusive process $\sigma^{\gamma\gamma} = \sigma(p p \to
H ) \times Br(H \to \gamma \gamma)$. This (\siggaga) could provide possibly
the best mode to search for light Higgs Boson in mass range 110-140 GeV. 
ATLAS \cite{atlas_diphoton} reported 95\% CL exclusion limit of 
$\sigma^{\gamma\gamma}/\sigma^{\gamma\gamma}_{SM} \sim 1.6 - 1.8$ in
mass range 110-130 GeV. CMS  \cite{cms_diphoton} on the other hand
reported the exclusion limit of  
$\sigma^{\gamma\gamma}/\sigma^{\gamma\gamma}_{SM} \sim 1.5 - 2$ in
mass range 110-140 GeV. 

Dark Matter (DM) and Electroweak Symmetry Breaking (EWSB) are one of
the most important areas of research in particle physics and
cosmology. One of the main goal of LHC is to discover the Higgs Boson and
hence provide the information about the EWSB mechanism. A DM particle
is expected to be a weakly interacting massive particle (WIMP) with
mass around EWSB scale. In SM the EWSB is achieved by a Higgs doublet
developing a vacuum expectation value (vev). Inert Higgs Doublet Model
(IHDM) is a very simple extension of the SM proposed by Deshpande and
Ma \cite{Deshpande:1977rw} to explain DM. IHDM is basically a two Higgs Doublet
Model with imposed $Z_2$ symmetry. IHDM due to the imposed $Z_2$
symmetry exhibits very interesting phenenomenology. It predicts the
existence of a heavy scalar field as a WIMP candidate. The rich
phenomenology of IHDM had been extensively discussed in the context of
DM phenomenology \cite{idm_dm,Dolle:2009fn}, neutrino mass \cite{Ma:2006km},
naturalness \cite{Barbieri:2006dq} and colliders
\cite{Lundstrom:2008ai,idm_lhc}. 

In this work we will analyze the
effect of IHDM on \htogaga in the light of recent results on the Higgs
Boson searches from LHC. This effect will mainly come from charged Higgs boson
 contributions as well as from the total decay width of the Higgs boson in case 
 that the invisible decay of the Higgs into dark matter is open. 
 We will show in this study that the IHDM 
 can account for the excess in the di-photon cross-section 
 reported  by ATLAS/CMS but it can also account for a deficit in 
 the di-photon cross-section without modifying the gluon
   fusion rate as well as  the other 
channels like $h\to b\bar{b}, \tau^+\tau^-, WW^*,ZZ^*$.

The paper is organized as follow;
In section \ref{section:2} we will give the
details of the IHDM. Section  \ref{section:3} is devoted 
to theoretical and experimental constraints while in 
\ref{section:4} we give detail of the evaluation of $h\to \gamma\gamma$
as well as phenomenological observable at LHC.
In section \ref{section:5} we will present our numerical analysis 
and finally we  conclude in section \ref{section:6}.  

\section{Inert Higgs Doublet Model \label{section:2}}

The Inert Higgs Doublet Model (IHDM) \cite{Deshpande:1977rw} is  an 
extension of the SM Higgs sector that could provide DM particles. 
Apart from the SM Higgs doublet $H_1$
it has an additional Higgs doublet $H_2$. In addition there is a $Z_2$
symmetry under which all the SM fields and $H_1$ are even while $H_2
\to - H_2$ under $Z_2$. We 
further assume that $Z_2$ symmetry is not spontaneously broken {\sl
i.e.} $H_2$ field does not develop vacuum expectation value
(vev). These doublets in terms of physical fields can be parameterized
as : 
\beq
H_1 =
   \left( \begin{array}{c}  \phi_1^+ \\ v/\sqrt{2} + 
(h + i \chi)/\sqrt{2}  \end{array}
     \right)  ~~~~, ~~
H_2 =    \left( \begin{array}{c}  \phi_2^+ \\ 
(S  + i A)/\sqrt{2}  \end{array}  \right)
\eeq
The $Z_2$ symmetry naturally imposes the flavor conservation.
The scalar potential allowed by $Z_2$ symmetry can be written as :
\beq
V = \mu_1^2 |H_1|^2 + \mu_2^2 |H_2|^2 + \lambda_1 |H_1|^4
+ \lambda_2 |H_2|^4 +  \lambda_3 |H_1|^2 |H_2|^2 + \lambda_4
|H_1^\dagger H_2|^2
+ \frac{\lambda_5}{2} \left\{ (H_1^\dagger H_2)^2 + h.c. \right\}
\label{potential}
\eeq
The electroweak gauge symmetry is broken by:
\beq
\langle H_1 \rangle = \left( \begin{array}{c} 0 \\ v/\sqrt{2} \end{array}
\right) ~~~~,~~
\langle H_2 \rangle = \left( \begin{array}{c} 0 \\ 0 \end{array}
\right)
\eeq
This pattern of symmetry breaking ensures unbroken $Z_2$ symmetry and
results in two CP even neutral scalars ($h$, $S$) one CP odd neutral
scalar ($A$) in addition to  a pair of charged scalars
($H^\pm$). There is no mixing between the two  
doublets and hence $h$ plays the role of the SM Higgs Boson. Note the
remaining Higgs Bosons namely $S, A$ and $H^\pm$ are ``inert'' and
they do not have any interaction with quarks and leptons. The $Z_2$
symmetry also ensures the stability of the lightest scalar ($S$ or
$A$) that can act 
as a dark matter candidate. This aspect has been extensively analyzed
in many works while exploring DM phenomenology of IHDM
\cite{idm_dm}. 
The masses of all these six scalars can be written in terms of six
parameters\footnote{$\mu_1^2$ is constrained by EWSB condition $v^2 = -
\mu_1^2/\lambda_1$}  namely
\beq
\{\mu_2^2, \lambda_1, \lambda_2, \lambda_3,\lambda_4,\lambda_5 \}
\label{eq:input_par_1}
\eeq
It is possible to write the quartic coupling $\lambda_i$ in terms of 
physical scalar masses and $\mu_2$ as follow:
\beqa
\lambda_1 &=& \frac{m_h^2}{2 v^2}\quad , \quad 
\lambda_3 = \frac{2}{v^2}\left(m_{H^\pm}^2 - \mu_2^2\right), \\
\lambda_4 &=& \frac{\left(m_S^2 + m_A^2 - 2 m_{H^\pm}^2\right)}{v^2} \quad ,
\quad 
\lambda_5 = \frac{\left(m_S^2 - m_A^2\right)}{v^2}
\label{lambds}
\eeqa
We are  then free to take as 6 independent parameters 
$(\lambda_i)_{i=1,\ldots , 5}$ and $\mu_2$
or equivalently the four physical scalar masses, $\lambda_2$ and $\mu_2$,
namely: 
\beq
\{\mu_2^2, m_h, m_S, m_A, m_{H^\pm}, \lambda_2 \}
\label{eq:input_par_2}
\eeq

\section{Theoretical and experimental constraints \label{section:3}}
  The parameter space of the scalar potential of the IHDM is reduced
both by theoretical constraints as well as by the results of
experimental searches. From the theoretical constraints which the IHDM
is subjected to, the most important are the ones that insure
tree-level unitarity and vacuum stability of the theory: 
%
%



\begin{itemize}
\item \underline{\bf Perturbativity :} We force the potential to be
  perturbative by requiring that all quartic couplings of the scalar potential 
Eq.~(\ref{potential}), obey
\beq
|\lambda_i| \le 8 \pi
\eeq
\item \underline{\bf Vacuum Stability :}
To get a potential $V$ bounded from below we obtain the following constraints
on the IDHM parameters:
\begin{eqnarray}
\lambda_{1,2} > 0 \quad \rm{and} \quad \lambda_3 + \lambda_4 -|\lambda_5| +
2\sqrt{\lambda_1 \lambda_2} >0 \quad\rm{and} \quad\lambda_3+2\sqrt{\lambda_1
  \lambda_2} > 0
\end{eqnarray}
\item\underline{\bf Unitarity :}
To constrain the scalar potential parameters of the
IHDM one can demand that tree-level unitarity is preserved
in a variety of scattering processes: scalar-scalar scattering,
gauge boson-gauge boson scattering and scalar-gauge boson scattering.
We will follow exactly the technique developed in 
\cite{Lee:1977eg} and therefore we limit ourselves to 
pure scalar scattering processes dominated by quartic interactions.

The full set of scalar scattering processes can be expressed 
as an $S$ matrix $ 22\times 22$ composed of 4 submatrices
 which do not couple with each other due to charge conservation and
CP-invariance \cite{unit1,abdesunit}. 
The entries are the quartic couplings which mediate
the scattering processes.

The eigenvalues are:
\begin{eqnarray}
&&e_{1,2}=\lambda_3 \pm \lambda_4 \quad , \quad
e_{3,4}= \lambda_3 \pm \lambda_5\\
&&e_{5,6}= \lambda_3+ 2 \lambda_4 \pm 3\lambda_5\quad , \quad
e_{7,8}=-\lambda_1 - \lambda_2 \pm \sqrt{(\lambda_1 - \lambda_2)^2 + \lambda_4^2}
\\
&& 
e_{9,10}= -3\lambda_1 - 3\lambda_2 \pm \sqrt{9(\lambda_1 - \lambda_2)^2 + (2\lambda_3 +
   \lambda_4)^2}
\\
&& 
e_{11,12}=
 -\lambda_1 - \lambda_2 \pm \sqrt{(\lambda_1 - \lambda_2)^2 + \lambda_5^2}
\end{eqnarray}
We impose perturbative unitarity constraint on all $e_i$'s. 
\beq
|e_i| \le 8 \pi ~, \forall ~ i = 1,...,12
\eeq 
the strongest constraint on $\lambda_{1,2}$ comes from $e_{9,10}$
which gives :
\beq
\lambda_1 + \lambda_2 \le \frac{8 \pi}{3} 
\eeq
%
%
\item\underline{\bf Electro Weak Precision Tests :} A common approach
to constrain physics beyond SM is using the global electroweak fit
through the oblique $S$, $T$ and $U$ parameters
\cite{Ref:PeskinTakeuchi}. 
In the SM the EWPT implies a relation between $m_h$ and $m_Z$. In this model,
there is also a relation among the masses. It follows from the expression for
$S$ and $ T$ that:
\begin{eqnarray}
 T &=& \frac{1}{32\pi^2 \alpha v^2}\Bigg[ F(m^2_{H^\pm}, m^2_{A})
+ F(m^2_{H^\pm}, m^2_{S}) - F(m^2_{A}, m^2_{S})\Bigg]
\end{eqnarray}
and,
\begin{eqnarray}
 S = \frac{1}{2\pi}\Bigg[ \frac{1}{6}\log(\frac{m^2_{S}}{m^2_{H^\pm}}) -
  \frac{5}{36} + \frac{m^2_{S} m^2_{A}}{3(m^2_{A}-m^2_{S})^2} + 
\frac{m^4_A (m^2_{A}-3m^2_{S})}{6(m^2_{A}-m^2_{S})^3} \log(\frac{m^2_{A}}{m^2_{S}})\Bigg]
\end{eqnarray}
where the function $F$ is defined by
\begin{equation}
	F(x,y) = \left\{ \begin{array}{lr}
		\frac{x+y}{2} - \frac{xy}{x-y}\log(\frac{x}{y}), & x\neq y \\
		0, & \, x=y
		\end{array}  \right.
\end{equation}
For the purpose of this paper, we will use the PDG values of
  $S$ and $T$ with $U$ fixed to be zero
  \cite{pdg,Kanemura:2011sj}. We allow $S$ and $T$ parameters to
be within 95\% CL. The central value of $S$ and
$T$, assuming a SM Higgs mass of $m_{H_{SM}} = 117$
GeV, are given by \cite{pdg} :
\beq
S = 0.03 \pm 0.09, ~ ~ T = 0.07 \pm 0.08
\eeq
with a fit correlation of 87\%. It appears that when the
  unitarity constraints and vacuum stabilities are applied, a bound on
  $m_{H^\pm}$ may be obtained. Note that we can restore custodial symmetry in
  the scalar potential of  IHDM by taking $m^2_{H^\pm} = m^2_{A}$.
%
%
\item\underline{\bf Experimental constraints:}
Here we will discuss the experimental constraints from direct searches 
on the masses of the IHDM. In the case of the SM Higgs ($h$), we can
use CMS and ATLAS constraints discussed in section \ref{section:1}
when the non SM Higgs decays such as $h\to SS$, $h\to H^+H^-$, $h\to
A^0A^0$ are kinematically forbidden.  
In the case where one of these 
decays is kinematically allowed, it will have a substantial branching ratio.
Therefore, it will suppress
the other SM decays and hence one can evade the present constraints on
SM Higgs which are based on conventional SM Higgs decays like $h\to
b\bar{b}$, $h\to \tau^+\tau^-$, $h\to WW^*$ and $h\to ZZ^*$. 
(see Fig.~\ref{fig:2}).\\
From the precise measurement of 
W and Z widths, one can get some constraints on the Higgs masses 
 by demanding that the decays $W^\pm\to \{SH^\pm, A^0H^\pm\}$ 
 and/or $Z\to \{SA^0,H^+H^-\}$ are forbidden. This leads to the following
 constraints: $m_S+m_{H^\pm}>m_W$, $m_A+m_{H^\pm}>m_W$, $m_A+m_S>m_Z$ and
$m_{H^\pm}>m_Z/2$ \cite{pdg}.\\
Additional constraints on the charged Higgs $H^\pm$ and CP-odd $A^0$ masses
can be derived. Note that LEP, Tevatron and LHC
 bounds on $H^\pm$ and $A^0$ can not apply because the standard 
search channels assumes that those scalars decays into a pair of fermions 
 which are absent in the IHDM due to $Z_2$ symmetry. 
In the IHDM, the charged Higgs $H^\pm$ can decay into $H^\pm\to W^\pm A^0$
followed by $A^0\to SZ$ or  $H^\pm\to W^\pm S$. Therefore the decay 
product of the production processes
$e^+e^-/pp\to H^+H^-$, $e^+e^-/pp\to SA^0$ 
would be missing energy and multi-leptons or multi-jets depending 
on the decay product of W and Z. Such signature would be similar to some
extent to the supersymmetric searches for charginos and neutralinos at 
$e^+e^-$ or at hadron colliders \cite{idm_lhc}. 
Taking into account those considerations, 
we will assume that $m_{H\pm}> 70$ GeV (see \cite{idm_lhc} for more details).
\end{itemize}

\section{$h\to\gamma\gamma$ in IHDM \label{section:4}}
It is well known that for the SM Higgs searches, the low mass 
$m_H\in [110,140]$ GeV, is the most challenging for LHC searches.
In this low mass region, the main search is through the 
 di-photons which can be complemented by the $\tau^+ \tau^-$ mode 
and potentially with the $b \bar{b}$ mode, while the $WW^*, ZZ^*$ channels
are already competitive in the upper edge ($130$--$140$ GeV) 
of this mass range \cite{ATLAS-CONF-2011-157-CMS-PAS-HIG-11-023}.

The SM predictions for the one-loop induced $h\to \gamma\gamma$
 have been worked out since many years \cite{smm}. 
It is  well known that the SM rate for $h\to \gamma\gamma$ 
 is dominated by the W loops and the branching ratio
 of this channel is of the order of $2\times 10^{-3}$.
Several studies have been carried out looking for
large loop effects beyond SM. Such large effects can be found in various 
extensions of the SM, such as the Minimal Supersymmetric Standard Model (MSSM)
\cite{mssm}, the Next-to-MSSM \cite{nmssm},
the two Higgs Doublet Model \cite{2hdm1,2hdm2,2hdm3}, 
the little Higgs models \cite{little}, 
extra-dimensions \cite{extra} and in models with triplet Higgs
\cite{triplet}. 

In the IHDM, the partial width of $h\to\gamma\gamma$ receives an additional
contribution from the charged Higgs boson loop which can both
  lower and raise the width compared to the SM. It can be expressed 
as \cite{abdel}:
\begin{eqnarray}
\Gamma(h\to \gamma\gamma) = \frac{\alpha^2 G_F
  m^2_{h}}{128\sqrt{2}\pi^3}\Bigg| \sum_{i} N_{ci} Q^2_{i} F_{i}  
+ g_{hH^{\pm}H^{\mp}}\frac{m^2_W}{m^2_{H^\pm}}F_{0}(\tau_{H^\pm}) \Bigg|^2 ,
\label{widths}
\end{eqnarray}
where $N_{ci}$, $Q_i$ are the color factor and the electric charge
respectively for a particle $i$ running in the loop. The dimensionless 
loop factors for particles of spin given in the subscript are:
\begin{eqnarray}
F_1 = 2+3\tau + 3\tau(2-\tau)f(\tau), \quad F_{1/2} =
-2\tau[1+(1-\tau)f(\tau)], \quad F_0 = \tau[1-\tau f(\tau)],
\end{eqnarray}
with
\begin{equation}
f(\tau) = \left\{ \begin{array}{lr}
[\sin^{-1}(1/\sqrt{\tau})]^2, & \tau \geq 1 \\
-\frac{1}{4} [\ln(\eta_+/\eta_-) - i \pi]^2, & \, \tau < 1
\end{array}  \right.
\end{equation}
and
\begin{equation}
\tau_i = 4 m_i^2 / m_h^2, \qquad \eta_{\pm} = 1 \pm \sqrt{1-\tau}.
\end{equation}
In Eq.~(\ref{widths}), the coupling $ g_{hH^{\pm}H^{\mp}}$ is given by
\begin{equation}
g_{hH^{\pm}H^{\mp}} = -2\,i\,\frac{m_W s_W}{e}\lambda_3 =  
-\,i\,\frac{e}{2s_W m_W} (m_{H\pm}^2-\mu_2^2)
\label{hgaga_coupl}
\end{equation}

It is clear from the above Eq.~(\ref{hgaga_coupl}), 
the coupling of the  SM higgs boson to a pair of charged 
Higgs is completely fixed by $\lambda_3$
parameter. As we will see later, the sign of $\lambda_3$ will play an important
role in the evaluation of the partial width of $h\to \gamma\gamma$.

More important than the branching ratios, however, is the total cross-section
of $\sigma^{\gamma\gamma}_{h}=\sigma(pp\to h \to \gamma \gamma)$, 
 since that is what is measured at the collider. 
The largest contribution to the production cross-section for this
observable $\sigma^{\gamma\gamma}_{h}$ is through gluon fusion, $gg \to h \to
\gamma \gamma$. 
For phenomenological purpose, we define the ratio of the di-photon cross
section normalized to SM rate as follow:
 \begin{eqnarray}
R_{\gamma\gamma}& =& 
\frac{\sigma_{h}^{\gamma\gamma}}{\sigma_{h_{SM}}^{\gamma\gamma}}=
\frac{\sigma(gg\to h)\times Br(h \to \gamma\gamma) }{\sigma(gg\to
  h)^{SM}\times Br(h \to \gamma\gamma)^{SM} } 
= \frac{Br(h \to \gamma\gamma) }{Br(h \to \gamma\gamma)^{SM} }
\label{ratio}
\end{eqnarray}
Where we have used the narrow width approximation in the first line of 
Eq.~(\ref{ratio}) while we have used the fact 
that $\sigma(gg\to h)$ is the same both in the IHDM and SM. 
One conclude that the ratio $R_{\gamma\gamma}$ in the IHDM depend 
only on the branching ratio of $h\to \gamma\gamma$. 
In the evaluation of the branching ratios, we use for the total decay widths
 the following expressions:
\begin{eqnarray}
\Gamma_h^{SM}&=&\sum_{f=\tau,b,c}\Gamma(h\to ff) + \Gamma(h\to WW^*)+
\Gamma(h\to ZZ^*) + \Gamma(h\to gg)+\Gamma(h\to \gamma\gamma)\\
\Gamma_h^{IHDM}&=&\Gamma_h^{SM}+\sum_{\Phi=S,A,H^\pm}\Gamma(h\to \Phi\Phi)
\end{eqnarray}                            
where the expressions for the scalar decay widths are taken from \cite{abdel}.
Note that the decays $h\to SS$, $h\to A^0A^0$ and $h\to H^\pm H^\mp$ might be
or not kinematically open. In the case where the Dark matter particle is
lighter than $m_h/2$, the decay $h\to SS$ is open and could give 
substantial contribution to
the total width of the Higgs. 
For future use, we give here the analytical expression for 
$hSS$ coupling in the IHDM: 
\begin{equation}
g_{hSS} = -2\,i\,\frac{m_W s_W}{e}\lambda_L =  
-\,i\,\frac{e}{s_W m_W} (m_{S}^2-\mu_2^2)
\label{hss_coupl}
\end{equation}
which is proportional to $(m_S^2-\mu_2^2)$, with 
 $\lambda_L=\lambda_3+\lambda_4 + \lambda_5$.
%
\section{Numerical results \label{section:5}}
\begin{figure}[htb]
\bcen
\epsfig{file=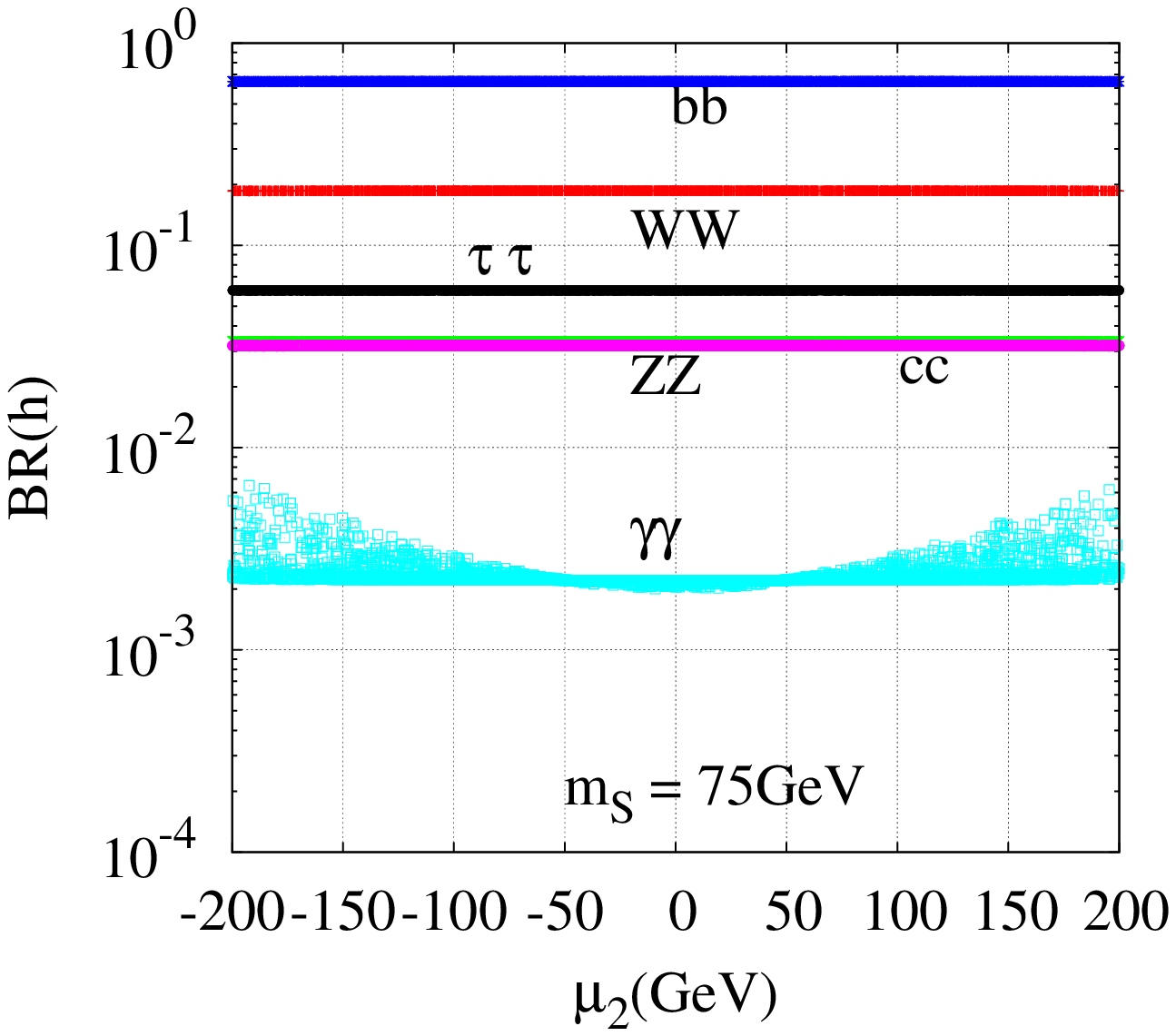,width=0.5\textwidth}\hspace*{-17cm}
\epsfig{file=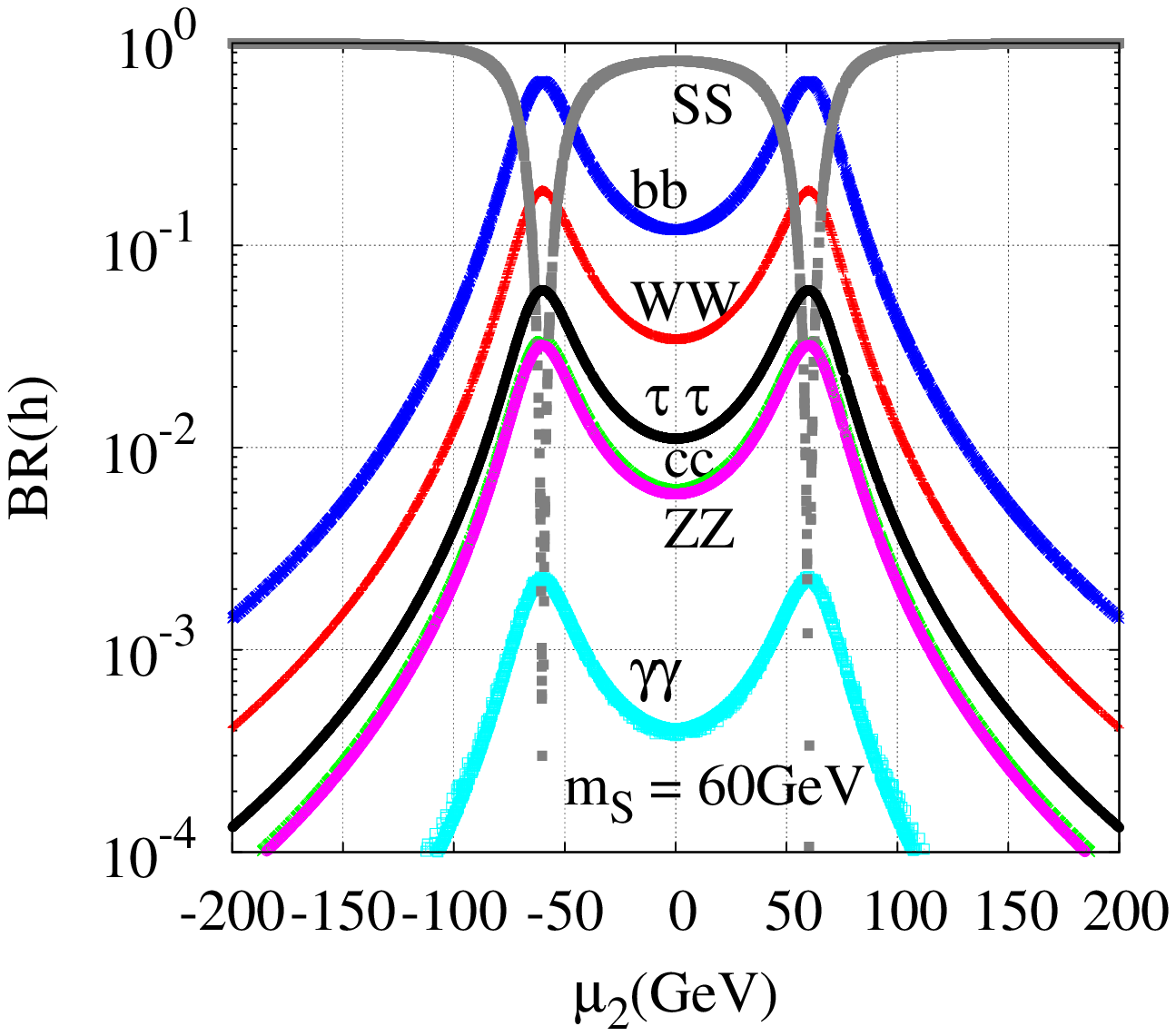,width=0.5\textwidth}
\caption{\sl Branching ratio of  Higgs boson $h$ as a function of $\mu_2$
(GeV) in the IHDM with $m_S=60$ GeV (left panel) and $m_S=75$ GeV
(right panel). We have chosen $m_h = 125$ GeV and have varied 
other parameters in range $70 < m_{H^\pm}, m_A < 1000$, $0 < \lambda_2 <
8 \pi$, $-500 < \mu_2 < 500$.}    
\label{fig:2}
\ecen
\end{figure}

Before presenting our numerical results we would like to point out that
in Ref. \cite{2hdm1} $h\to \gamma\gamma$ has been studied in 2HDM type I 
 as well as in the IHDM. But, Ref. \cite{2hdm1} 
 only focused on the parameter space region where only SM decays
namely $h \to \tau^+ \tau^-, b \bar{b}, c \bar{c},  W^+ W^-, Z Z,
\gamma\gamma, g g $ decays of the SM Higgs are kinematically
allowed. In this case, the total width of the Higgs boson is the same 
in SM and in IHDM and therefore our ratio $R_{\gamma\gamma}$ given in Eq.~
(\ref{ratio}) reduce to: $\Gamma(h\to \gamma\gamma)/\Gamma(h\to
\gamma\gamma)^{SM}$ defined in Ref. \cite{2hdm1}. Our results agree
with the results given in \cite{2hdm1}. In the case where  
$h\to SS$ is open, the ratio $\Gamma(h\to \gamma\gamma)/\Gamma(h\to
\gamma\gamma)^{SM}$ is not the appropriate one to be compared with CMS and
ATLAS data but rather the ratio of branching ratio as defined in 
Eq.~(\ref{ratio}). Moreover, as we will show, 
in our study we discuss the effect of all parameters that have some 
impact on the ratio $R_{\gamma\gamma}$ like charged Higgs mass, 
dark matter particle mass as well as the sign of the coupling $hH^{\pm}H^{\mp}$. We
will also comment on constraints coming from WIMP relic density.\\
As shown in Fig.~\ref{fig:2} we can have a very interesting
situation of evading the LHC bounds on SM Higgs in the case where the invisible
decay of Higgs ($h \to S S$) is kinematically allowed. This issue will be
discussed in details  in forthcoming work \cite{progress}. 

\begin{figure}[htb]
\bcen
\vspace*{-.3cm}
\hspace*{-1.5cm}
\epsfig{file=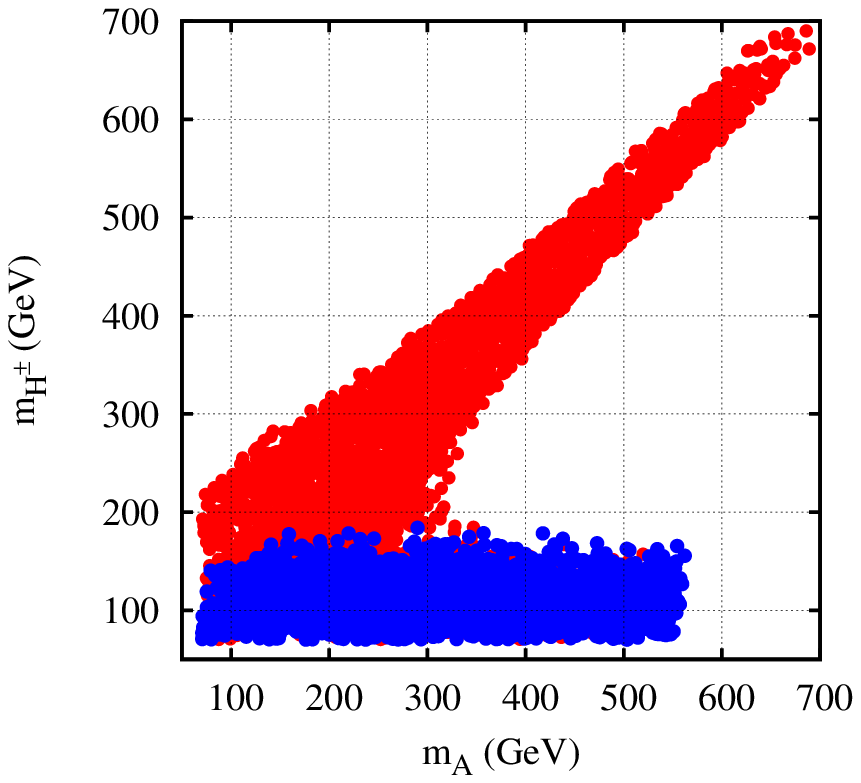,width=0.7\textwidth}\hspace*{-4cm}
\epsfig{file=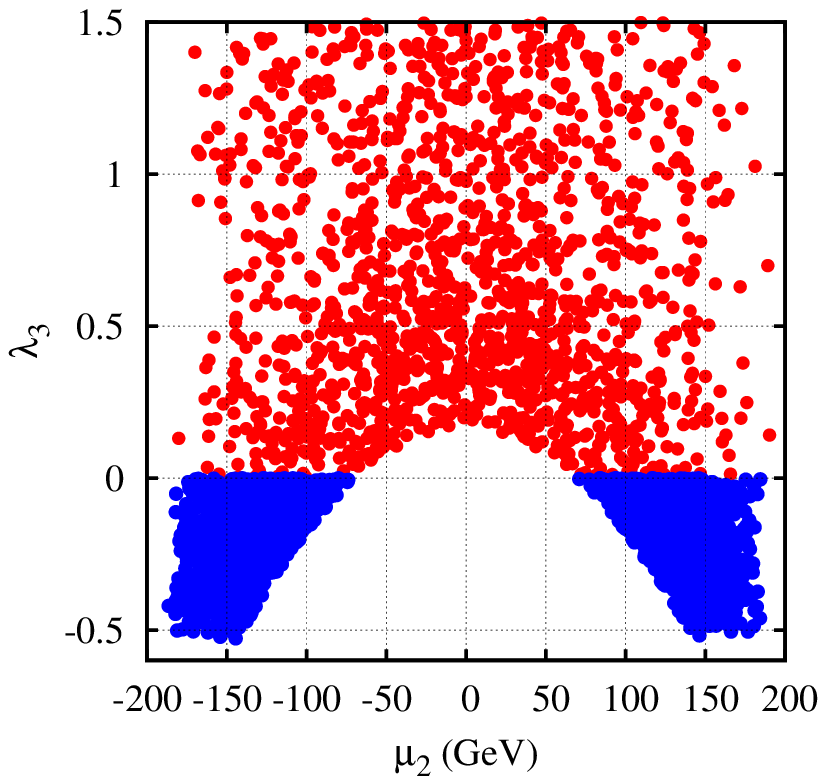,width=0.7\textwidth}
\vspace*{-1.5cm}
\caption{\sl The allowed parameter space in the $(m_A,m_{H^\pm})$
plane (left panel) and $(m_\mu, \lambda_2)$ plane (right panel) taking into
account theoretical and experimental constraints. The red dots
(\redbullet) represent $R_{\gamma\gamma} < 1$ and blue dots
(\bluebullet) represent $R_{\gamma\gamma} > 1$.} 
\label{fig:1}
\ecen
\end{figure}

\begin{figure}[htb]
\bcen
\vspace*{-.3cm}
\epsfig{file=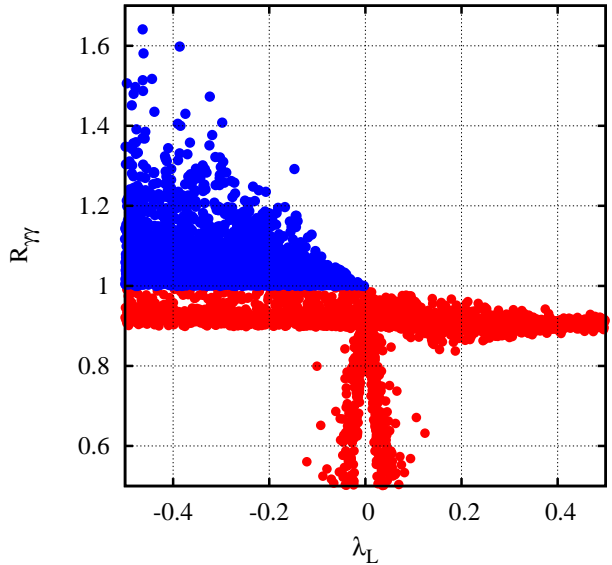,width=0.7\textwidth}
\vspace*{-1cm}
\caption{\sl \rgaga as a function of $\lambda_L$ with the range of
parameters as given in Eq.~(\ref{eq:para_scan}).  }  
\label{addfig}
\ecen
\end{figure}

\begin{figure}[htb]
\bcen
\hspace*{-1.9cm}
\epsfig{file=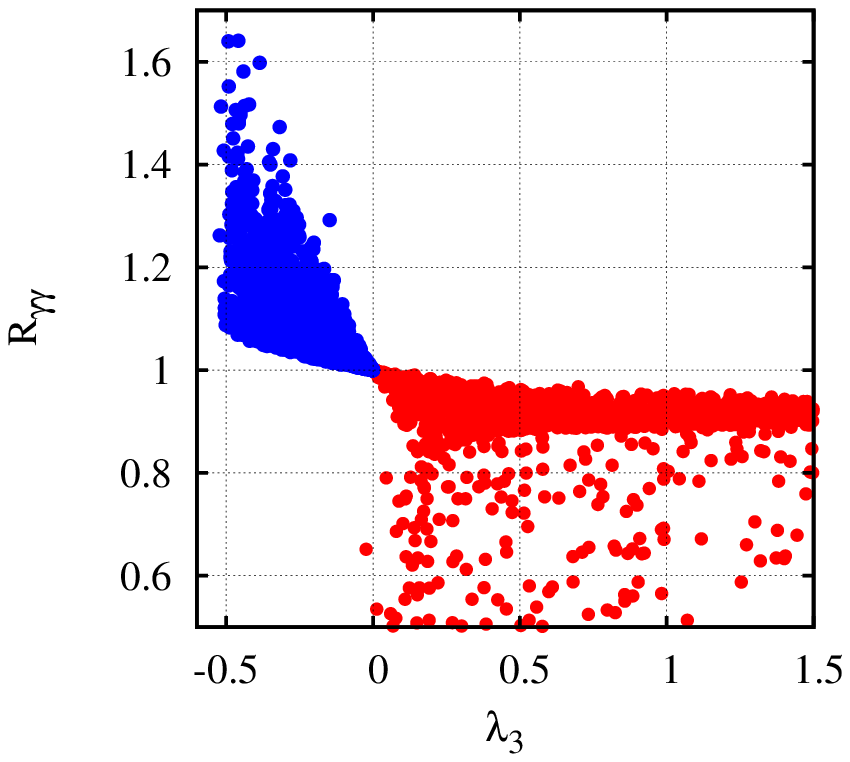,width=0.7\textwidth}\hspace*{-4cm}
\epsfig{file=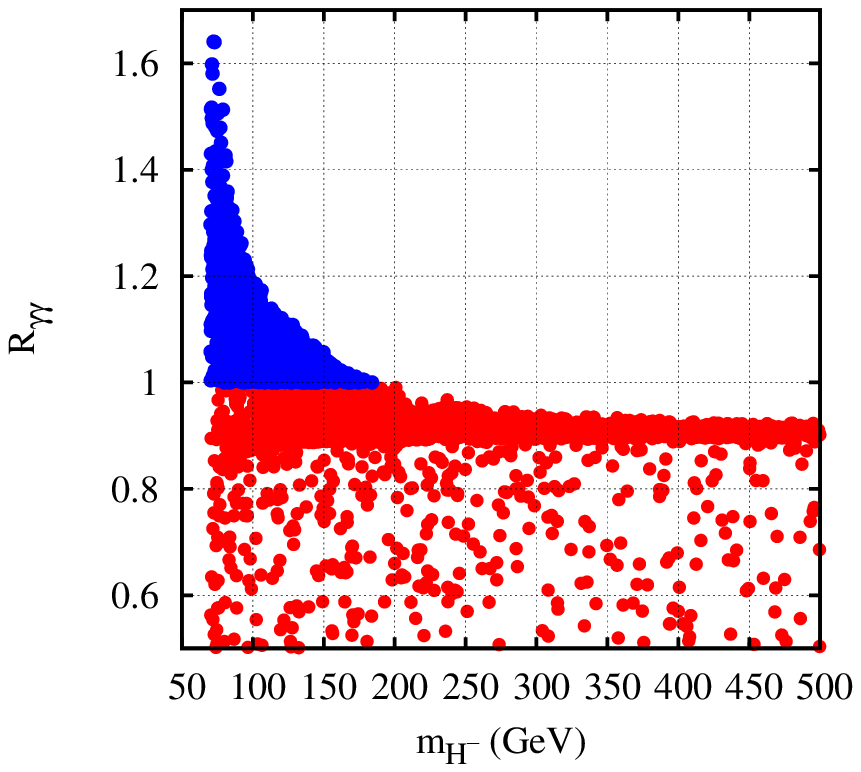,width=0.7\textwidth}
\vspace*{-1.5cm}
\caption{\sl Range of values of $R_{\gamma\gamma}$ accessible in the IHD model as a
function of $\lambda_3$ (left) and $m_{H^\pm}$. }
\label{fig:3}
\ecen
\end{figure}

\begin{figure}[htb]
\bcen
\hspace*{-1.7cm}
\epsfig{file=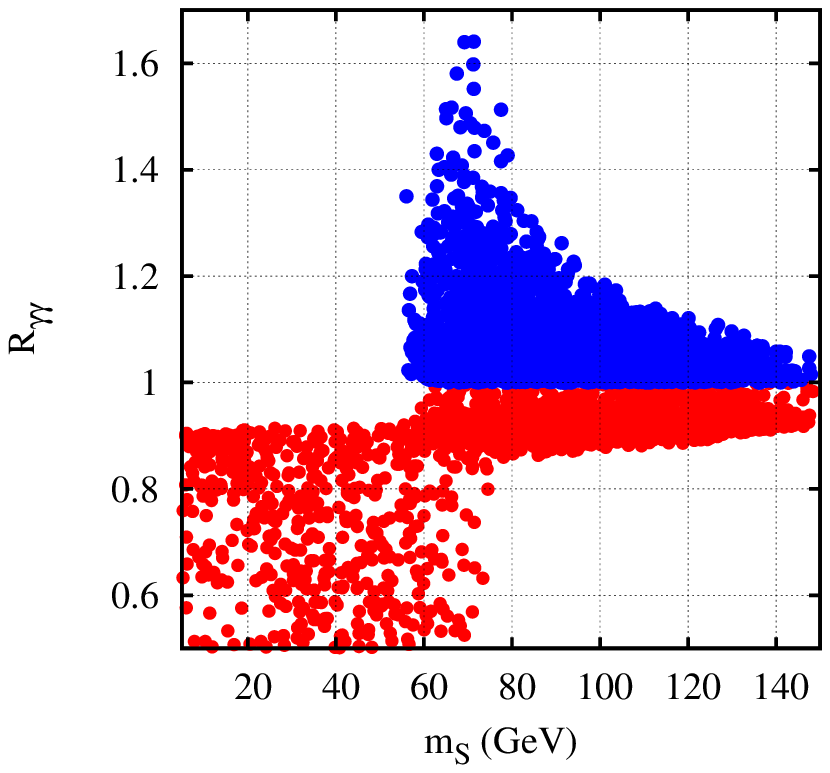,width=0.74\textwidth}\hspace*{-4cm}
\epsfig{file=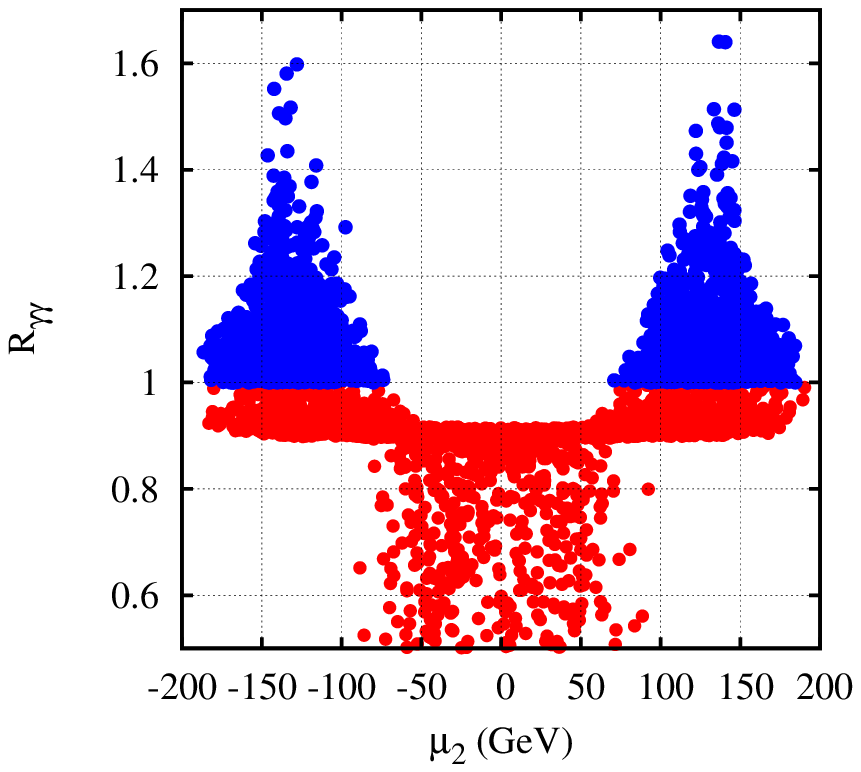,width=0.74\textwidth}
\vspace*{-1.5cm}
\caption{\sl Range of values of $R_{\gamma\gamma}$ accessible in the IHD model as a
function of $\lambda_3$ (left) and $m_{H^\pm}$. The parameter space are the
same as in Figure \ref{fig:1}.}
\label{fig:4}
\ecen
\end{figure}

In our numerical analysis, we perform a systematic scan over the 
parameter space of the IHDM. We vary the IHDM parameters in the range 
\begin{eqnarray}
   &  110\, \rm{GeV} \le  m_h \le 150\, \rm{GeV} \nonumber \\
    & 5\, \rm{GeV} \le m_{S} \le 150\, \rm{GeV}&  \nonumber \\ 
    & 70\, \rm{GeV} \le m_{H^\pm} , m_A \le 1000\,\rm{GeV}, & \nonumber \\
    & -500\, \rm{GeV} \le \mu_{2} \le 500\, \rm{GeV}, & \nonumber \\
    &  0\, \le \lambda_2 \le 8\pi &  
\label{eq:para_scan}
\end{eqnarray}
In addition we have imposed $m_S < m_A$ and $m_S < m_{H^\pm} $ and $m_S <
m_h$. This mass hierarchy ensures that $m_S$ could be the WIMP DM
candidate. 
These values cover essentially the entire physically interesting range of
parameters in the IHDM. For SM Higgs ($h$) we have specifically chosen
a range where $h\to\gamma\gamma$ could be a important channel (light Higgs Boson
mass) and the region that shows some deviations from SM as reported in
recent LHC results \cite{atlas_diphoton},\cite{cms_diphoton}. 
We have imposed the theoretical constraints mentioned above as well as 
constraints form oblique parameters $S$ and $T$.\\
In addition, we would like to stress in passing that the coupling $hSS$, which 
is proportional to $\lambda_L=\lambda_3+\lambda_4+\lambda_5$ 
(see Eq.~(\ref{hss_coupl})), is an important piece for the calculation of the 
 WIMP relic density \cite{Dolle:2009fn}. It has been show in 
\cite{Dolle:2009fn}, that with light Higgs boson $m_h\sim 120$ GeV 
and $m_S\sim 60$--$80$ GeV a relic density would be consistent with
experimental data for $|\lambda_L|<0.2$. In the following numerical analysis,
 with low Higgs mass 115--140 GeV, we will impose rather conservative 
limit on $|\lambda_L|<0.5$.
With all the above constraints discussed, we get the following limits:
$\lambda_2< 4\pi/3$, $m_{H\pm}, m_A<700$ GeV and $|\mu_2|<200$ GeV.  
In all plots, the coding color are: 
red (\redbullet) and blue (\bluebullet) dots in scatter
plots shown in Figures \ref{fig:1}, \ref{addfig}, \ref{fig:3},\ref{fig:4} represents
$R_{\gamma\gamma} < 1$ and $R_{\gamma\gamma} > 1$ respectively. 

In Figure.~\ref{fig:1}, we show the allowed region in the $(m_A, m_{H^\pm})$
(left panel) and $(m_S, m_{H^\pm})$ (right panel).
The perturbativity and vacuum stability constraints
together dramatically reduce the allowed parameter space of the model. In
particular, the perturbativity and vacuum stability constraints excludes a
large values of charged Higgs mass $m_{H^\pm}$ and CP-odd mass $m_A$ while 
 EWPT measurement constraint mainly the splitting between the scalar masses. 
Accordingly, an enhancement in \rgaga is possible  for
relatively light charged Higgs mass. \\
In the right panel of Figure.~\ref{fig:1} we shown the scatter plot in
$(\mu_2, \lambda_3)$ space. As can be seen again from this figure the
enhancement in \rgaga is possible only for negative values of
$\lambda_3$. Note that the plots are symmetric under $\mu_2 \to -\mu_2$.

In Fig.~(\ref{addfig}), we illustrate \rgaga as a function of $|\lambda_L|<0.5$
which is the main parameter contributing to the WIMP relic density
calculation. For large and negative $\lambda_L$ one can see that 
\rgaga can reach 1.6 while for large and positive $\lambda_L$,  
\rgaga can be of the order 0.7.
It is clear from this plot that even for small $|\lambda_L|<0.2$, 
which might be needed to accommodate WIMP relic density \cite{Dolle:2009fn}, 
we still have both cases with   \rgaga $<1$ and $>1$.

As discussed in section.\ref{section:1} due the presence of 
$Z_2$ symmetry in IHDM the lightest Higgs boson will be a stable particle. 
With the spectrum we have chosen the lightest $Z_2$ odd Higgs 
Boson is the neutral scalar $S$ and hence SM Higgs will be having 
an invisible decay mode namely $h\to S S$. For illustration, 
in Figure.~\ref{fig:2}, we have fixed $m_h = 125$ GeV 
and shown the branching ratios as a function of $\mu_2$ for $m_S=60$ GeV (left
plot) and $m_S=75$ GeV (right plot). 
In the left panel of Figure.~\ref{fig:2}, with $m_S = 60$ GeV, 
the invisible decay 
$h\to SS$ is open and dominate over all other SM decays except 
around $\mu_2\sim m_S$  where the coupling $hSS$ which is proportional to
$(m_S^2-\mu_2^2)$ (Eq.~\ref{hss_coupl}) 
gets suppressed and hence the situation becomes 
similar to  SM.  In the case where $h\to SS$ dominate, 
the partial width of $h\to SS$ contribute significantly to the total width of
the Higgs which becomes substantial resulting in a suppression of the 
$Br(h\to \gamma\gamma)$ which is always smaller than its SM value. 
For $|\mu_2| \sim m_S$, the $Br(h\to \gamma\gamma)$ can
reach the full SM value.
We can observe from Figure. \ref{fig:2} (left) that the branching fraction
of the invisible decay of SM Higgs ($h \to S S$) could be very large 
 resulting on a suppression of the other modes such as $b\bar{b}$, 
$WW$, $ZZ$ and $\tau^+\tau^-$ and
hence one can evade the present experimental constraints on the SM
Higgs mass based on $WW$, $ZZ$ and $\tau^+\tau^-$. 
We will discuss this in future work \cite{progress}. The invisible
decay of SM Higgs could evade some of the constraints on SM Higgs
Boson this has been extensively studied in many phenomenological
studies \cite{invisible_higgs}.   

In the right panel of Figure.~\ref{fig:2}, we take $m_S = 75$ GeV and 
then the decay $h\to SS$ is close. In this case, the total decay 
width of the Higgs boson is similar in both SM and IHDM. 
Therefore, the partial decay width $\Gamma(h\to \gamma\gamma)$ 
will receive only smooth variation through the charged Higgs contribution. 
So that an enhancement of the branching
ratio $Br(h\to \gamma\gamma)$ with up to a factor of 2 over the SM is possible. 
In fact, in our parameterization of IHDM given in Eq.~(\ref{eq:input_par_2}), 
$\lambda_3$ is fixed by the charged Higgs mass and $\mu_2$ 
parameter through Eq.~(\ref{lambds}). The sign of $\lambda_3$
is then completely fixed by the sign of $m^2_{H^\pm} - \mu^2_2$. 
Hence, for small $|\mu_2| < m_{H^\pm}$, the sign of $\lambda_3$ is 
positive. In this case, the charged Higgs contribution to 
$\Gamma(h\to \gamma\gamma)$ is totally destructive with the SM. While
for large $|\mu_2| > m_{H^\pm}$, $\lambda_3$ becomes negative and the
charged Higgs contribution to $\Gamma(h\to \gamma\gamma)$ 
becomes constructive with SM and  gets substantial enhancement.



In Fig. \ref{fig:3} (left panel) we have shown $R_{\gamma\gamma}$ as a
function of  $\lambda_3$. The other parameters are taken as specified in 
Eq.~(\ref{eq:para_scan}).
As can be seen from the left plot, as seen previously, 
IHDM can  increase the value of \rgaga ($>1$) only for negative
values of $\lambda_3$ where the charged Higgs contribution is constructive
with the W loops. For positive $\lambda_3$, the charged Higgs contribution
is destructive with the W loops resulting in a suppression of $h\to
\gamma\gamma$ rate. \\
The dependence of $R_{\gamma\gamma}$ on
 the charged Higgs mass is illustrated in the right panel of Figure.\ref{fig:3}.
The variation of $R_{\gamma\gamma}$ as a function of $m_{H\pm}$ 
scales almost like $1/m^2_{H^\pm}$.
Varying $m_{H^\pm}$ between 70 GeV and 190 GeV results in dramatic change of 
$R_{\gamma\gamma}$ from 1.5 to 1. We stress that even for light charged 
Higgs $m_{H\pm}\in [70,190]$ GeV, we could have $R_{\gamma\gamma}<1$.
This could be due to the possible opening of 
the invisible decay $h\to SS$ which could reduce significantly the branching
fraction of $h\to \gamma\gamma$ or to the fact that $\mu_2$ is rather small
making $\lambda_3$ positive.\\
Note that if we relax the constraint on $\lambda_L$ discussed above, 
we can get large $\lambda_3$ in the following range: $\lambda_3\in [-1.5,2]$. 
A large and negative $\lambda_3\sim -1.5$ would give a constructive 
charged Higgs contribution with the W loops and therefore 
amplify \rgaga which can reach 1.6--2.2 for light charged Higgs 
$\sim 70-100$ GeV.

In Fig.~\ref{fig:4}, we show $R_{\gamma\gamma}$ as a function of $m_S$ and
$\mu_2$. From Figure.~\ref{fig:4} (left panel) one can observe that 
 an enhancement compared to SM value in \rgaga is only possible 
for $m_S > m_h/2$  while for $m_S<m_h/2$ a suppression of 
\rgaga is guaranteed.  
Similarly one can observe that \rgaga can
be enhanced with respect to SM value for relatively 
large value of $\mu_2$ while for small $|\mu_2|<70$ GeV 
\rgaga is suppressed. 


\section{Conclusions\label{section:6}}
To summarize, in this work we study $h\to \gamma\gamma$ in the 
IHDM by imposing vacuum stability, perturbativity, unitarity and
precision electroweak measurements. We have shown that within the allowed
range of the IHDM model $h\to\gamma\gamma$ could show substantial deviation from
the SM result. Hence the observation of Higgs boson in $h\to\gamma\gamma$ could 
help us in constraining the parameter space of the model. We have also
shown that observation of \rgaga $> 1$ or $< 1$ could rule out a large
portion of the allowed parameter space of IHDM. \\
Taking into account all the constraints defined 
in section \ref{section:3} there is an upper
bound on $m_{H^\pm}$ and $m_{A}$ as evident from Figure \ref{fig:1}
(left panel). This bound essentially comes from Unitarity of the model. 
If the CMS and ATLAS excess in the di-photon channel 
is confirmed with more data, having \rgaga $>1$ would 
 favor the following scenarios:
\begin{itemize}
\item $\lambda_3<0$, i.e $|\mu_2| > m_{H^\pm}$ 
\item Charged Higgs Boson mass ($m_{H^\pm}$) will be bounded
($\lesssim 200$ GeV).    
\end{itemize}
On the other hand, if with more data we have
$R_{\gamma\gamma}<1$ this  
scenario would favor either a light DM particle $m_S<m_h/2$ 
such that $h\to SS$ is open and/or a positive $\lambda_3$ {\sl i.e.} 
$m_{H^\pm} > |\mu_2|$. 

\section*{Acknowledgements}
%
We would like to thank Chuan-Hung Chen and Gilbert Moultaka
 for useful discussions.
A.A would like to thank NSC-Taiwan for partial support during his stay 
 at NCKU where part of this work has been done.
The work of R.B was supported by the Spanish Consejo Superior de
Investigaciones Cientificas (CSIC). 
The work of N.G. is supported by grants from Department of Science \& 
Technology (DST), India under project number SR/S2/HEP-09/10 and
University Grants Commission (UGC), India under project number
38-58/2009(SR). 


\end{document}